\documentclass[aps,pre,twocolumn,showpacs,amsmath,amssymb,superscriptaddress]{revtex4}

\usepackage{graphicx}
\graphicspath{{pict/}}
\usepackage{dcolumn}
\usepackage{bm}

\begin{document}

\preprint{APS/123-QED}

\title{Discrete Nonlinear Schr{\"o}dinger Equations Free of the Peierls-Nabarro Potential}

\author{S. V. Dmitriev}
\affiliation{Institute of Industrial Science, University of
Tokyo, 4-6-1 Komaba, Meguro-ku, Tokyo 153-8505, Japan \\
$^2$ Department of Mathematics and Statistics, University of
Massachusetts Lederle Graduate Research Tower,
Amherst, MA 01003-4515, USA}

\author{P. G. Kevrekidis}
\affiliation{Department of Mathematics and Statistics, University of
Massachusetts Lederle Graduate Research Tower,
Amherst, MA 01003-4515, USA}

\author{A. A. Sukhorukov}
\affiliation{Nonlinear Physics Centre, Research School of Physical Sciences and Engineering, \\
Australian National University, Canberra ACT 0200, Australia}

\author{N. Yoshikawa}
\affiliation{Institute of Industrial Science, University of
Tokyo, 4-6-1 Komaba, Meguro-ku, Tokyo 153-8505, Japan \\
$^2$ Department of Mathematics and Statistics, University of
Massachusetts Lederle Graduate Research Tower,
Amherst, MA 01003-4515, USA}

\author{S. Takeno}
\affiliation{Graduate School, Nagasaki Institute of Applied Science,
Nagasaki 851-0193, Japan}

\date{\today}

\begin{abstract}
We derive a class of discrete nonlinear Schr{\"o}dinger (DNLS)
equations for general polynomial nonlinearity whose stationary
solutions can be found from a reduced two-point algebraic problem.
It is demonstrated that the derived class of discretizations
contains subclasses conserving classical norm or a modified norm
and classical momentum. These equations are interesting from the
physical standpoint since they support stationary discrete
solitons free of the Peierls-Nabarro potential. As a consequence,
even in highly-discrete regimes, solitons are not trapped by the
lattice and they can be accelerated by even weak external fields.
Focusing on the cubic nonlinearity we then consider a small
perturbation around stationary soliton solutions and, solving
corresponding eigenvalue problem, we (i) demonstrate that solitons
are stable; (ii) show that they have two additional zero-frequency
modes responsible for their effective translational invariance;
(iii) derive semi-analytical solutions for discrete solitons
moving at slow speed. To highlight the unusual properties of
solitons in the new discrete models we compare them with that of
the classical DNLS equation giving several numerical examples.
\end{abstract}


\maketitle

\section{Introduction}

Recent theoretical and experimental results have demonstrated that the fundamental properties of waves can be precisely engineered by introducing a periodic modulation of the medium characteristics. In particular, there appear unique possibilities to control the propagation of light in photonic structures with a periodic modulation of the optical refractive index~\cite{Joannopoulos:1995:PhotonicCrystals}, and manage the flow of atomic Bose-Einstein condensates (BEC) in periodic potentials~\cite{Eiermann:2003-60402:PRL}. Additional flexibility in tailoring wave dynamics can be realized through nonlinear self-action, that may appear due to various physical mechanisms such as light-matter interactions for optical beams or atom-atom scattering in BEC. One of the key effects of nonlinearity is the suppression of the natural tendencies of localized wave packets to broaden due to dispersion or diffraction, supporting the formation of optical lattice solitons~\cite{Christodoulides:2003-817:NAT, Kivshar:2003:OpticalSolitons} and self-localized atomic states~\cite{Trombettoni:2001-2353:PRL, Eiermann:2004-230401:PRL}.

The excitation of lattice solitons and their dynamics can be described by the discrete nonlinear Schr\"odinger (DNLS) equation when the wave propagation is primarily defined by tunneling between neighboring potential wells~\cite{Christodoulides:1988-794:OL, Christodoulides:2003-817:NAT}. This special regime of energy flow results in special properties of discrete solitons. It was predicted theoretically~\cite{Kivshar:1993-3077:PRE, Aceves:1996-1172:PRE, Bang:1996-1105:OL, Krolikowski:1996-876:JOSB,papa} and observed experimentally~\cite{Morandotti:1999-2726:PRL, Fratalocchi:2005-1808:OE} that the discrete solitons can freely propagate through the lattice below a certain energy threshold, whereas at higher energies the solitons become trapped at a particular lattice site. This phenomenon is related to the presence of the self-induced Peierls-Nabarro potential (PNp) barrier, defining the difference of energies between the solitons whose centers are positioned at a lattice site or in-between the neighboring lattice sites.

The ability to enhance or suppress PNp may allow for precise
control over transmission of higher-energy wave packets through
the lattice. The PNp vanishes for discrete solitons in the
framework of the Ablowitz-Ladik (AL)
equations~\cite{Ablowitz:1975-598:JMP, Ablowitz:1976-1011:JMP}
which are integrable. One of the features of the AL model is
nonlocal nonlinearity, as the medium response is defined by the
wave amplitudes at the neighboring lattice sites, in contrast to
the local on-site nonlinearity that leads to strong
PNp~\cite{Kivshar:1993-3077:PRE}. However, the AL model has not
been directly connected to a specific physical application. On the
other hand, more recently, it was demonstrated that nonlocal
nonlinearity of a more general type representing real physical
systems can indeed lead to strong reduction of
PNp~\cite{Oster:2003-56606:PRE, Xu:2005-113901:PRL} for lattice
solitons. It was also shown that PNp can be reduced (and even
reversed) in the case of a saturable nonlinear
response~\cite{Hadzievski:2004-33901:PRL}.

In this paper, we consider a class of DNLS equations featuring a general type of nonlinearity characterized by a nonlocal response as well as an arbitrary polynomial dependence on the intensity, that includes the case of nonlocal saturable nonlinearity. We show that, under certain conditions, the PNp can be made exactly zero. Quite remarkably, the corresponding equations are generally not integrable, leading to nontrivial soliton dynamics.
We present examples of equations that conserve the total energy (or norm), suggesting that the predicted effects can be observed in specially engineered periodic structures.

Our presentation is structured as follows. In section~\ref{sec:setup}, we
present the general setup of the continuum and discrete models of
the present work. In section~\ref{sec:pnpFree}, we derive the class of PNp-free
DNLS equations and extract two subclasses conserving different
physical quantities. The case of cubic (Kerr) nonlinearity is
considered in detail in section~\ref{sec:Kerr}. Finally, in section~\ref{sec:Conclusions}, we briefly summarize our findings and present our conclusions.

\section{Setup} \label{sec:setup}

The PNp potential is a feature of discrete models and does not appear for continuous and translationally invariant nonlinear Schr\"odinger (NLS) equation. We demonstrate below that, by performing appropriate discretization of the continuous NLS equation, it is possible to obtain a broad class of DNLS models where PNp is also absent. We note that a conceptually similar approach has been developed
for Klein-Gordon type lattices~\cite{Speight:1994-475:NLN,
Speight:1999-1373:NLN, Kevrekidis:2003-68:PD, Dmitriev:2005-7617:JPA,
Dmitriev:nlin.PS/0506002:ARXIV, Barashenkov:2005-35602:PRE,
Oxtoby:2006-217:NLN, Speight:nlin.PS/0509047:ARXIV}
and is now starting to emerge in NLS settings as well
\cite{Kevrekidis:submitted}. We also note in passing the interesting
concurrent efforts but in a different direction (namely that of obtaining
exact solutions with a free parameter and demonstrating the absence
for those of PN barriers) of \cite{sax1,sax2,sax3}.
Here, we present our methodology for the generalized NLS equation of the form
\begin{eqnarray} \label{NLSE}
   i\psi_t+\frac{1}{2} \psi_{xx}+G^{\prime}(|\psi|^2)\psi =0,
\end{eqnarray}
where $\psi(x,t)$ is a complex function of two real variables;
$G(\xi)$ is a real function of its argument and
$G^{\prime}(\xi)=dG/d\xi$.

We introduce the lattice sites $x_n=nh$, where $h$ is the lattice
spacing and $n=0,\pm 1,\pm 2,...$  We also introduce the
following shorthand notations
\begin{eqnarray} \label{Notations}
  \psi(x_{n-1}) = \psi_{-}, \quad
  \psi(x_{n}) = \psi_{n},\quad
  \psi(x_{n+1})=\psi_{+},
\end{eqnarray}
and will focus only on discretizations that involve such nearest
neighbor sites.

Our more specific aim will be
to construct the discrete analogues of Eq. (\ref{NLSE})
of the form:
\begin{eqnarray} \label{deq1}
   i \dot{\psi}_n
   + r( \psi_{-}, \psi_{-}^{\star}, \psi_n, \psi_n^{\star},
        \psi_{+}, \psi_{+}^{\star} )
   = 0,
\end{eqnarray}
such that the ansatz
\begin{eqnarray} \label{AnsatzDiscrete}
   \psi_n(t)=f_n e^{i\omega t},
\end{eqnarray}
reduces Eq. (\ref{deq1}) to the three-point discrete problem of
the form
\begin{eqnarray} \label{Threepoint}
   -\omega f_n + R(f_{-},f_{n},f_{+}) = 0,
\end{eqnarray}
whose solution can be found from a reduced two-point discrete
problem $u(f_{-},f_{n}) = 0$. Such a selection will entail a
mono-parametric freedom for the resulting algebraic problem
leading to stationary state solutions. This will, in turn, be
responsible for the effective translational invariance in what
follows. Such models will be called the models free of
Peierls-Nabarro potential (PNp-free models, for short) since their
stationary solutions can be placed anywhere with respect to the
lattice.

Following the method proposed in \cite{Kevrekidis:submitted}, we derive a wide
class of PNp-free DNLS equations. Then we demonstrate that they
contain subclasses conserving the classical norm
\begin{eqnarray}
   N=\sum_n |\psi_n|^2, \label{Norm}
\end{eqnarray}
or modified norm
\begin{eqnarray} \label{newnorm}
   \tilde{N} = \frac{1}{2} \sum_{n=-\infty}^{\infty}
              \left( \psi_n \psi_{+}^{\star}
                     + \psi_n^{\star} \psi_{+} \right),
\end{eqnarray}
and classical momentum
\begin{eqnarray} \label{MomNLSE}
   P = i \sum_{n=-\infty}^{\infty}
        \left( \psi_n \psi_{+}^{\star} -
          \psi_n^{\star} \psi_{+} \right) .
\end{eqnarray}

\section{PNp-free DNLS equations} \label{sec:pnpFree}
\subsection{Auxiliary problem}

Firstly, we formulate an auxiliary problem. Seeking stationary
solutions of Eq. (\ref{NLSE}) in the form
\begin{eqnarray} \label{Ansatz}
  \psi(x,t) = f(x)e^{i\omega t},
\end{eqnarray}
we reduce it to an ordinary differential equation for the real
function $f(x)$,
\begin{eqnarray} \label{KGstatic}
   D(x)\equiv f^{\prime \prime} - 2\omega f + 2fG^{\prime}(f^2) = 0,
\end{eqnarray}
having the first integral
\begin{eqnarray} \label{KGstaticFI}
   u(x)\equiv \left(f^{\prime}\right)^2 - 2\omega f^2 + 2G(f^2) + C =
   0,
\end{eqnarray}
where $C$ is the integration constant.

We then identify discretizations of Eq. (\ref{KGstatic}) of the form
\begin{eqnarray}
   D(f_{-},f_n,f_{+}) = 0, \label{KGstaticDiscrete}
\end{eqnarray}
such that solutions to the three-point discrete Eq.
(\ref{KGstaticDiscrete}) can be found from a reduced two-point
problem
\begin{eqnarray}
   u(f_{-},f_n)\equiv \frac{1}{h^2}\left(f_n-f_{-}\right)^2 \nonumber
   \\ - 2\omega f_{-}f_n + 2G(f_{-},f_n) + C = 0,
   \label{ReducedDiscrete}
\end{eqnarray}
which is a discrete version of Eq. (\ref{KGstaticFI}), assuming
that $G(f_{-},f_n)$ reduces to $G(f^2)$ in the continuum limit
($h\rightarrow 0$). In the present study, we will set $C=0$, which
is sufficient for obtaining the single-humped stationary
solutions. However, the case of arbitrary $C$ enables one to
construct {\em all} possible stationary solutions to the
corresponding discrete model \cite{Dmitriev:submitted}.

Taking into account that Eq. (\ref{KGstatic}) is the static
Klein-Gordon equation with the potential
\begin{eqnarray}
   V(f)=\omega f^2 - G(f^2), \label{Potential}
\end{eqnarray}
a wide class of discretizations solving the auxiliary problem has
been offered in the very recent work of \cite{Dmitriev:2005-7617:JPA, Dmitriev:nlin.PS/0506002:ARXIV}.

For example, discretizing the left-hand side of the identity
$(1/2)du/df=D(x)$, we obtain the discrete version of Eq.
(\ref{KGstatic}),
\begin{eqnarray}
   D_1(f_{-},f_n,f_{+})\equiv
   \frac{u(f_n,f_{+})-u(f_{-},f_n)}{f_{+}-f_{-}}=0. \label{D1}
\end{eqnarray}
Formally, $D_1(f_{-},f_n,f_{+})=0$ is a three-point problem but,
clearly, its solutions can be found from the two-point problem
$u(f_{-},f_n)=0$ and thus, the auxiliary problem is solved. We note,
in passing,
that this type of argument was first proposed in
\cite{Kevrekidis:2003-68:PD}.

Before we come back to our main problem of finding PNp-free
discretizations for Eq. (\ref{NLSE}), we should remark that among
the solutions to the auxiliary problem we should select the ones
which can be rewritten in terms of $\psi_n$ and $\psi_n^{\star}$
in the desired form of Eq. (\ref{deq1}). This can be done easily,
e.g., when $D_1$ given by Eq. (\ref{D1}) is written in a
non-singular form (i.e., when the denominator cancels with an
appropriate factoring of the numerator).  This always occurs if
$G(\xi)$ is polynomial and if $u(f_{-},f_n)$ possesses the
symmetry
\begin{eqnarray}
   u(f_{-},f_n)=u(f_n,f_{-}). \label{Symmetry}
\end{eqnarray}

We thus focus on $G(\xi)$ in the form of Taylor expansion,
\begin{eqnarray}
   G(|\psi|^2)=\sum_{k=1}^{\infty}a_k|\psi|^{2k}, \label{Gpolynomial}
\end{eqnarray}
with real coefficients $a_k$.

The most general polynomial, two-point discrete version of Eq.
(\ref{Gpolynomial}), possessing the symmetry
$G(f_{-},f_n)=G(f_n,f_{-})$ is
\begin{eqnarray}
   G(f_{-},f_n)=\sum_{k=1}^{\infty}a_k Q_{2k}(f_{-},f_n),
   \label{GSymmetric}
\end{eqnarray}
where $Q_{2k}(f_{-},f_n)$ involves only the terms of order $2k$:
\begin{equation}
   Q_{2k}(f_{-},f_n)=\frac{\sum_{i=0}^{k}
   q_{i,k}\left(f_{-}^{i}f_n^{2k-i}+f_{-}^{2k-i}f_n^{i}\right)}
   {2\sum_{i=0}^{k}q_{i,k}},\label{Qk}
\end{equation}
with $q_{i,k}$ being free parameters such that
$\sum_{i=0}^{k}q_{i,k} \neq 0$.

Then we substitute Eq. (\ref{ReducedDiscrete}) with $G(f_{-},f_n)$
given by Eq. (\ref{GSymmetric}) into Eq. (\ref{D1}) to obtain
\begin{eqnarray}
   -\omega f_n + \frac{1}{2h^2}\left( f_{-} - 2f_n + f_{+}
   \right)\nonumber \\ +\frac{1}{2}\sum_{k=1}^{\infty} \frac{a_k
   S_{2k-1}(f_{-},f_n,f_{+})}{\sum_{i=0}^{k}q_{i,k}}=0,
   \label{D1final}
\end{eqnarray}
where the function
\begin{eqnarray}
   S_{2k-1}(f_{-},f_n,f_{+})= q_{0,k}
   \sum_{m=0}^{2k-1}f_{-}^{m}f_{+}^{2k-1-m}
   \nonumber \\
      +\sum_{i=1}^{k} q_{i,k} \Big(f_n^i \sum_{m=0}^{2k-i-1}
       f_{-}^{m}f_{+}^{2k-i-1-m}
   \nonumber
   \\+f_n^{2k-i}\sum_{m=0}^{i-1}f_{-}^{m}f_{+}^{i-1-m} \Big),\label{Sk}
\end{eqnarray}
involves only the terms of order $2k-1$.

\subsection{Main problem} \label{SubSecMainProblem}

PNp-free discretization of NLS equation Eq. (\ref{NLSE}) has the
form:
\begin{eqnarray}
   i \dot{\psi}_n + \frac{1}{2h^2}\left( \psi_{-} - 2\psi_n +
   \psi_{+} \right) \nonumber \\
   +\frac{1}{2}\sum_{k=1}^{\infty} \frac{a_k {\cal S}_{2k-1}
   (\psi_{-},\psi_n,\psi_{+})}{\sum_{i=0}^{k}q_{i,k}}=0,
   \label{DNLSE}
\end{eqnarray}
where ${\cal S}_{2k-1} (\psi_{-},\psi_n,\psi_{+})$ is any function
that, upon substituting Eq. (\ref{AnsatzDiscrete}), reduces to
$S_{2k-1} (f_{-},f_n,f_{+})e^{i\omega t}$, with
$S_{2k-1}(f_{-},f_n,f_{+})$ given by Eq. (\ref{Sk}). Indeed, the
stationary solutions to Eq. (\ref{DNLSE}), satisfying the
three-point problem, Eqs. (\ref{D1final}) and (\ref{Sk}), can be
found from the reduced two-point problem, Eqs.
(\ref{ReducedDiscrete}), (\ref{GSymmetric}), and (\ref{Qk}).

To complete the construction of the PNp-free DNLS equation we need to
obtain a suitable function ${\cal S}_{2k-1} (\psi_{-}, \psi_n,
\psi_{+})$ reducible by the ansatz Eq. (\ref{AnsatzDiscrete}) to
$S_{2k-1}(f_{-},f_n,f_{+}) e^{i\omega t}$. Clearly,
$S_{2k-1}(f_{-}, f_n, f_{+})$ can have several counterparts ${\cal
S}_{2k-1} (\psi_{-}, \psi_n, \psi_{+})$. For example, both
$|\psi_n|^2\psi_{-}$ and $\psi_n^2\psi_{-}^{\star}$ give
$(f_n^2f_{-})e^{i\omega t}$ upon substituting Eq.
(\ref{AnsatzDiscrete}).

As it can be seen from Eq. (\ref{Sk}), the typical term of $S_{2k-1}
(f_{-}, f_n, f_{+})$ is $f_{-}^k f_{n}^l f_{+}^m$ with
$k+l+m=2k-1$, and $k,l,m \ge 0$. This term can be transformed,
e.g., to
\begin{eqnarray}
   f_{-}^k f_{n}^l f_{+}^m \rightarrow \psi_{-}|\psi_{-}^{k-1}
   \psi_{n}^l \psi_{+}^m|, \nonumber \\ \psi_{n}|\psi_{-}^{k}
   \psi_{n}^{l-1} \psi_{+}^m|, \nonumber \\
   \psi_{+}|\psi_{-}^{k} \psi_{n}^l \psi_{+}^{m-1}|,
   \label{Transformations}
\end{eqnarray}
for $k>0$, $l>0$, and $m>0$, correspondingly.

It can also be transformed, e.g., to
\begin{eqnarray}
   f_{-}^k f_{n}^l f_{+}^m \rightarrow
   \psi_{-}^{\star}\psi_{n}^2|\psi_{-}^{k-1} \psi_{n}^{l-2}
   \psi_{+}^m|, \label{Transformations1}
\end{eqnarray}
for $k>0$ and $l>1$, or to
\begin{eqnarray}
   f_{-}^k f_{n}^l f_{+}^m \rightarrow
   \left(\psi_{-}^2\right)^{\star}\psi_{n}^3|\psi_{-}^{k-2}
   \psi_{n}^{l-3} \psi_{+}^m|, \label{Transformations2}
\end{eqnarray}
for $k>1$ and $l>2$, and so on.

One can see that the number of possibilities rapidly increases
with increase in the order of the term, $2k-1$.

\subsection{PNp-free DNLS equation conserving $N$}

Let us consider the following PNp-free DNLS equation
\begin{eqnarray}
   i \dot{\psi}_n + \frac{1}{2h^2}\left( \psi_{-} - 2\psi_n +
   \psi_{+} \right) \nonumber \\
   +\frac{1}{2}\sum_{k=1}^{\infty} \frac{a_k {\cal S}_{2k-1}
   (\psi_{-},\psi_n,\psi_{+})}{\sum_{i=1}^{k}q_{i,k}}=0,
   \label{NormConsDNLSE}
\end{eqnarray}
where
\begin{eqnarray}
   {\cal S}_{2k-1} (\psi_{-},\psi_n,\psi_{+})
   \nonumber \\
   =\psi_n\sum_{i=1}^{k}q_{i,k}\Big(\Big| \psi_n^{i-1}\Big|
   \sum_{m=0}^{2k-i-1}
   \Big|\psi_{-}^{m} \psi_{+}^{2k-i-1-m}\Big| \nonumber \\
   +\Big| \psi_n^{2k-i-1}\Big|\sum_{m=0}^{i-1} \Big|\psi_{-}^{m}
   \psi_{+}^{i-1-m} \Big| \Big).\label{SkNormCons}
\end{eqnarray}
For this model, the classical norm of Eq.
(\ref{Norm}), is conserved.

To construct this equation we had to set $q_{0,k}=0$ to drop the
first sum in the right-hand side of Eq. (\ref{Sk}), because it
does not contain $f_n$.

It is possible to construct some other DNLS equations conserving
$N$ and one example will be given later for the Kerr nonlinearity.
However, the above equation is interesting because it involves
only the {\em on-site} nonlinearities modified through inter-site
coupling.

Stationary solutions to Eq. (\ref{NormConsDNLSE}) can be found
from the two point problem Eq. (\ref{ReducedDiscrete}) with
$G(f_{-},f_n)$ given by Eq. (\ref{GSymmetric}), where
\begin{equation}
Q_{2k}(f_{-},f_n)=\frac{\sum_{i=1}^{k}
q_{i,k}\left(f_{-}^{i}f_n^{2k-i}+f_{-}^{2k-i}f_n^{i}\right)}
{2\sum_{i=1}^{k}q_{i,k}},\label{QkNormCons}
\end{equation}
i.e., in comparison with Eq. (\ref{Qk}), we simply drop the term
with $i=0$.

\subsection{PNp-free DNLS equation conserving $\tilde{N}$ and $P$}

In Eq. (\ref{Qk}), let us drop all the terms corresponding to odd
$i$ . We can write the result in the form
\begin{equation}
   Q_{2k}(f_{-},f_n)=\frac{\sum_{i=0}^{\lfloor k/2 \rfloor}
   q_{2i,k}\left(f_{-}^{2i}f_n^{2k-2i}+f_{-}^{2k-2i}f_n^{2i}\right)}
   {2\sum_{i=0}^{\lfloor k/2 \rfloor}q_{2i,k}},\label{QkMomtCons}
\end{equation}
where $\lfloor k/2 \rfloor$ is the largest integer not greater
than $k/2$. Substituting Eqs. (\ref{ReducedDiscrete}),
(\ref{GSymmetric}), and (\ref{QkMomtCons}) into Eq. (\ref{D1}) we
obtain
\begin{eqnarray}
   -\omega f_n + \frac{1}{2h^2}\left( f_{-} - 2f_n + f_{+}
   \right)\nonumber \\ +\frac{1}{2}\sum_{k=1}^{\infty} \frac{a_k
   S_{2k-1}(f_{-},f_n,f_{+})}{\sum_{i=0}^{\lfloor k/2
   \rfloor}q_{2i,k}}=0, \label{D1finalMomtCons}
\end{eqnarray}
where
\begin{eqnarray}
   S_{2k-1}(f_{-},f_n,f_{+})\nonumber \\
   = (f_{-} + f_{+}) \Big\{ q_{0,k} \sum_{m=0}^{k-1} f_{-}^{2m}
   f_{+}^{2(k-m-1)}
   \nonumber \\
   +\sum_{i=1}^{\lfloor k/2
      \rfloor} q_{2i,k} \Big(f_n^{2i}
       \sum_{m=0}^{k-i-1}f_{-}^{2m}f_{+}^{2(k-i-m-1)}
   \nonumber
   \\+f_n^{2(k-i)}
        \sum_{m=0}^{i-1}f_{-}^{2m}f_{+}^{2(i-m-1)} \Big)
        \Big\} .\label{SkMomtCons}
\end{eqnarray}

The resulting dynamical model (corresponding to
(\ref{D1finalMomtCons}) and Eq. (\ref{SkMomtCons})) of the form
\begin{eqnarray}
   i \dot{\psi}_n + \frac{1}{2h^2}\left( \psi_{-} - 2\psi_n +
   \psi_{+} \right) \nonumber \\
   +\frac{1}{2}\sum_{k=1}^{\infty} \frac{a_k {\cal S}_{2k-1}
   (\psi_{-},\psi_n,\psi_{+})}{\sum_{i=0}^{\lfloor k/2
   \rfloor}q_{2i,k}}=0, \label{MomtConsDNLSE}
\end{eqnarray}
with
\begin{eqnarray}
   {\cal S}_{2k-1}(\psi_{-},\psi_n,\psi_{+})\nonumber \\
   = (\psi_{-} + \psi_{+}) \Big\{ q_{0,k} \sum_{m=0}^{k-1}
   \Big|\psi_{-}^{2m} \psi_{+}^{2(k-m-1)}\Big|
   \nonumber \\
   +\sum_{i=1}^{\lfloor k/2
   \rfloor}q_{2i,k}\Big(\Big|\psi_n^{2i}\Big|\sum_{m=0}^{k-i-1} \Big|
   \psi_{-}^{2m}
   \psi_{+}^{2(k-i-m-1)}\Big| \nonumber \\
   +\Big|\psi_n^{2(k-i)}\Big| \sum_{m=0}^{i-1} \Big|\psi_{-}^{2m}
   \psi_{+}^{2(i-m-1)}\Big| \Big) \Big\}, \label{SkMomtCons1}
\end{eqnarray}
respects the conservation laws of both the modified norm of
Eq. (\ref{newnorm}) and the momentum of Eq. (\ref{MomNLSE}).
Stationary solutions to Eq. (\ref{MomtConsDNLSE}) can be found
from the two-point problem Eqs. (\ref{ReducedDiscrete}),
(\ref{GSymmetric}), and (\ref{QkMomtCons}).

\section{Cubic (Kerr) nonlinearity} \label{sec:Kerr}

\subsection{Examples of PNp-free models} \label{ExamplesOfPNpFree}

Let us study in detail Eq. (\ref{NLSE}) with Kerr nonlinearity,
i.e., with $G^{\prime}(|\psi|^2)=|\psi|^2$,
\begin{eqnarray}
   i\psi_t+\frac{1}{2} \psi_{xx}+|\psi|^2\psi =0. \label{NLSEKerr}
\end{eqnarray}
We then have $G(|\psi|^2)=|\psi|^4/2$, i.e., in Eq.
(\ref{Gpolynomial}), $a_2=1/2$ and $a_k=0$ for $k \neq 2$.

To construct PNp-free DNLS equations we write Eq. (\ref{D1final})
and Eq. (\ref{Sk}) for the case of Kerr nonlinearity as
\begin{equation}
   -\omega f_n + \frac{1}{2h^2}\left( f_{-} - 2f_n + f_{+}
   \right)
   + \frac{ S_{3} (f_{-},f_n,f_{+}) }{4(\alpha+\beta+\gamma)}=0,
   \label{Kerr1}
\end{equation}
and
\begin{eqnarray}
   S_{3}(f_{-},f_n,f_{+})
   = \alpha(f_{+}^{3}+f_{-}f_{+}^2+f_{-}^{2}f_{+}+f_{-}^{3})
   \nonumber \\
   +\beta(f_{+}^{2}+f_{-}f_{+}+f_{-}^{2}+f_n^2)f_n
   \nonumber \\
   +2\gamma(f_{+}+f_{-})f_n^2,\label{Sk2Kerr}
\end{eqnarray}
where we introduced the following shorter notations for the free
parameters: $\alpha=q_{0,2}$, $\beta=q_{1,2}$, and
$\gamma=q_{2,2}$.

Solutions to the three-point problem of Eq. (\ref{Kerr1}) can be
found from the following two-point problem [Eqs.
(\ref{ReducedDiscrete}), (\ref{GSymmetric}), and (\ref{Qk})]
\begin{eqnarray}
   \frac{1}{h^2}\left(f_n-f_{-}\right)^2-2\omega f_{-}f_n \nonumber\\
   + \frac{\alpha(f_{-}^4+f_n^4) + \beta(f_{-}f_n^3+f_{-}^3f_n) +
   2\gamma f_{-}^2f_n^2 }{2(\alpha+\beta+\gamma)} = 0.
   \label{ReducedKerr}
\end{eqnarray}

The ensuing PNp-free DNLS equation with Kerr nonlinearity is
\begin{equation}
   i \dot{\psi}_n + \frac{1}{2h^2}\left( \psi_{-} - 2\psi_n +
   \psi_{+} \right) +\frac{ {\cal S}_{3} (\psi_{-},\psi_n,\psi_{+})
   }{4(\alpha+\beta+\gamma)}=0, \label{DNLSEKerr}
\end{equation}
where ${\cal S}_{3} (\psi_{-},\psi_n,\psi_{+})$ is any function
that, upon substituting Eq. (\ref{AnsatzDiscrete}), reduces to
$e^{i\omega t}S_{3} (f_{-},f_n,f_{+})$ [i.e., respecting the
phase invariance of the equation], with
$S_{3}(f_{-},f_n,f_{+})$ given by Eq. (\ref{Sk2Kerr}). Some
guidelines on how to construct ${\cal S}_{3}
(\psi_{-},\psi_n,\psi_{+})$ can be found in Sec.
\ref{SubSecMainProblem}.

From Eq. (\ref{NormConsDNLSE}) and Eq. (\ref{SkNormCons}), at
$a_2=1/2$ and $a_k=0$ for $k\neq 2$, we obtain the {\em Kerr-type
DNLS equation conserving the classical norm $N$}
\begin{eqnarray}
   i \dot{\psi}_n + \frac{1}{2h^2}\left( \psi_{-} - 2\psi_n +
   \psi_{+} \right)\nonumber \\
   + \frac{ \psi_n }{4(\beta+\gamma)}\Big[ \beta\left( |\psi_{-}|^2
   +|\psi_{-}\psi_{+}|+|\psi_{+}|^2+|\psi_{n}|^2\right) \nonumber \\
   +2\gamma\left(|\psi_{-}\psi_{n}|+|\psi_{n}\psi_{+}|\right)\Big]=0.
   \label{Kerr1NormCons}
\end{eqnarray}
The two-point equation for finding the amplitudes of stationary
solutions to Eq. (\ref{Kerr1NormCons}) is Eq. (\ref{ReducedKerr})
with $\alpha=0$.
%

Similarly, from Eq. (\ref{MomtConsDNLSE}) and Eq.
(\ref{SkMomtCons1}), we obtain the {\em Kerr-type DNLS equation
conserving modified norm $\tilde{N}$ and momentum $P$}
\begin{eqnarray}
   i \dot{\psi}_n + \frac{1}{2h^2}\left( \psi_{-} - 2\psi_n +
   \psi_{+} \right)\nonumber \\
   + \frac{ \psi_{-}+\psi_{+} }{4(\alpha+\gamma)}\Big[ \alpha\left(
   |\psi_{-}|^2 + |\psi_{+}|^2\right) +2\gamma  |\psi_{n}|^2
   \Big]=0. \label{Kerr1MomtConserv}
\end{eqnarray}
Amplitudes of stationary solutions of Eq. (\ref{Kerr1MomtConserv})
satisfy Eq. (\ref{ReducedKerr}) with $\beta=0$.
%

Notice that the integrable discretization
of \cite{Ablowitz:1975-598:JMP, Ablowitz:1976-1011:JMP} is obtained
from Eq. (\ref{Kerr1MomtConserv}) as the special case of
$\alpha=0$. For $\alpha \neq 0$, this model can be regarded as a
Salerno-type model \cite{Salerno:1992-6856:PRA}, i.e., a homotopic continuation
including the integrable limit and reducing to NLS equation in the
continuum limit.

DNLS equations of the form of (\ref{Kerr1NormCons}) and
(\ref{Kerr1MomtConserv}) do not, of course, exhaust the list of
possible PNp-free models with Kerr nonlinearity. To give one more
example, we note that the last term of Eq. (\ref{Sk2Kerr}) can be
used to produce the following DNLS equation conserving classical
norm
\begin{eqnarray}
   i \dot{\psi}_n + \frac{1}{2h^2}\left( \psi_{-} - 2\psi_n +
   \psi_{+} \right) 
   +\frac{1}{4}\left( \psi_{-} + \psi_{+} \right)|\psi_n|^2 \nonumber \\
   +\frac{1}{4}\left( \psi_{-}^{\star} + \psi_{+}^{\star}
   \right)\psi_n^2 =0. \label{DNLSE11}
\end{eqnarray}
Amplitudes of stationary solutions to Eq. (\ref{DNLSE11}) can be
found from Eq. (\ref{ReducedKerr}) at $\alpha=\beta=0$.

\subsection{Soliton solutions} \label{SolitonSolutions}

We now compare some properties of the classical DNLS equation,
\begin{eqnarray}
   {\rm model}\,\, I:\,\,\,\,\,\,\,\,\,\,\,\,\,\,\,\,\,\,
\,\,\,\,\,\,\,\,\,\,\,\,\,\,\,\,\,\,\,\,\, \nonumber \\
   i\dot{\psi}_n + \frac{1}{2h^2}( \psi_{-} - 2\psi_n +
   \psi_{+})+|\psi_n|^2\psi_n=0,\label{Model1}
\end{eqnarray}
with these of the $N$-conserving model of Eq. (\ref{Kerr1NormCons}) with
$\beta=0$,
\begin{eqnarray}
{\rm model}\,\, II:\,\,\,\,\,\,\,\,\,\,\,\,\,\,\,\,\,\,\nonumber \\
i \dot{\psi}_n + \frac{1}{2h^2}\left( \psi_{-} - 2\psi_n +
\psi_{+} \right) \nonumber \\
+ \frac{ \psi_n }{2} \left( |\psi_{-}\psi_{n}| +|\psi_{n}\psi_{+}|
\right)=0;\label{Model2}
\end{eqnarray}
and those of the $\tilde{N}$- and $P$-conserving model of Eq.
(\ref{Kerr1MomtConserv}) with $\gamma=0$,
\begin{eqnarray}
   {\rm model}\,\, III:\,\,\,\,\,\,\,\,\,\,\,\,\,\,\,\,\,\,\,\nonumber \\
   i \dot{\psi}_n + \frac{1}{2h^2}\left( \psi_{-} - 2\psi_n +
   \psi_{+} \right)\nonumber \\
   + \frac{ \psi_{-}+\psi_{+} }{4}\left( |\psi_{-}|^2 +
   |\psi_{+}|^2\right) =0.\label{Model3}
\end{eqnarray}

All three models share the same continuum limit, the integrable
NLS equation Eq. (\ref{NLSEKerr}), and thus, in the regime of weak
discreteness (small lattice spacing $h$), their soliton solutions
of the form of Eq. (\ref{AnsatzDiscrete}) can be expressed {\it
approximately} as
\begin{eqnarray}
   \psi_{n}(t)=\frac{A}{\cosh[A h (n-x_0) ]}\exp[-i(A^2/2)t],
   \label{ApproxSOLITON}
\end{eqnarray}
where $A$ and $\omega=A^2/2$ are the soliton amplitude and
frequency, respectively.

The approximate solution of Eq. (\ref{ApproxSOLITON}) contains the
free parameter $x_0$ defining the soliton position. However, in
contrast to the NLS equation of Eq. (\ref{NLSEKerr}), where $x_0$
can be chosen arbitrarily due to translational invariance, the
DNLS models {\it usually} have stationary soliton solutions only
for a discrete set of values of $x_0$ (e.g. on-site, $x_0=0$, and
inter-site, $x_0=1/2$). This is true, for example, for the
classical DNLS of model $I$ and for the Salerno model
\cite{Salerno:1992-6856:PRA}, among others. The models $II$ and $III$, by
construction, are among the members of a wider class of DNLS
equations proposed in this paper, where stationary soliton
solutions exist for any $x_0$, or, in other words, they can be
placed anywhere with respect to the lattice; otherwise put, the
Peierls-Nabarro potential is absent for stationary solutions of
these models.

Let us now describe the exact soliton solutions to the models $I$,
$II$, and $III$.

An explicit formula does not exist for the stationary soliton
solutions of model $I$. Such solutions can be obtained using the
fixed point algorithms \cite{Kevrekidis:2001-2833:IJMPB}, but, as mentioned above, only
for $x_0=0$ and $x_0=1/2$.

Model $II$, model of Eq. (\ref{DNLSE11}), and also model of Eq.
(\ref{Kerr1MomtConserv}) at $\alpha=0$, have the same equations
for the amplitudes of stationary solutions. However, the latter
model, as mentioned above, is the integrable DNLS equation
\cite{Ablowitz:1975-598:JMP, Ablowitz:1976-1011:JMP} and thus, an exact stationary solutions for these models
can be obtained explicitly in the form
\begin{eqnarray}
   \psi_{n}(t)=\frac{1}{h}\frac{\sinh \mu}{\cosh[\mu (n-x_0)]}
   \exp^{i\omega t}, \label{ALsoliton}
\end{eqnarray}
where $x_0$ is the parameter defining the soliton position and it
can obtain any value from $[0,1)$. The soliton frequency
$\omega=h^{-2}(1-\cosh\mu)$ and amplitude $A=h^{-2}\sinh^2\mu$ are
expressed in terms of the free parameter $\mu>0$.

Model $III$ has the solutions of the form of Eq.
(\ref{AnsatzDiscrete}) with $f_n$ derivable from the
two-point problem
\begin{eqnarray}
   \frac{1}{h^2}\left(f_n-f_{-}\right)^2 - 2\omega f_{-}f_n +
   \frac{1} {2}(f_{-}^4+f_{n}^4) = 0.
   \label{ReducedDiscreteKerrMomtCut}
\end{eqnarray}
The soliton can be constructed by setting an arbitrary value for
$f_{-}$ (or $f_{n}$) in the range $[A_m,A_s]$ and finding $f_{n}$
(or $f_{-}$) from the quartic Eq.
(\ref{ReducedDiscreteKerrMomtCut}). Quantities $A_m$ and $A_s$ are
the amplitudes of solitons centered between two lattice sites and
on a lattice site, respectively. We have $A_m=\sqrt{2\omega}$, and
$A_s$ can be found from the condition that two distinct real roots
of Eq. (\ref{ReducedDiscreteKerrMomtCut}) merge into a multiple
root. The arbitrariness in the choice of initial value of $f_{-}$
(or $f_{n}$) implies the absence of the Peierls-Nabarro potential
and the possibility to place the soliton anywhere with respect to
the lattice.

\subsection{Soliton's internal modes} \label{InternalModes}

Let us study the stability of stationary soliton solutions for the
models described in Sec. \ref{SolitonSolutions}. In this study we
calculate the soliton's internal modes and frequencies of these
modes.

Following the methodology of the paper \cite{Carr:1985-201:PLA}, to study the
stability of the solution Eq. (\ref{AnsatzDiscrete}), we consider
the complex perturbation $\epsilon_n(t)$ in the frame rotating with
the periodic solution:
\begin{eqnarray}
   \psi_n(t)=\left[f_n+\epsilon_n(t)\right] e^{i\omega t}.
   \label{Perturbation}
\end{eqnarray}

Substituting Eq. (\ref{Perturbation}) into the classical DNLS equation
Eq. (\ref{Model1}) (model $I$) we find that the linearized
equation satisfied by $\epsilon_n(t)$ is
\begin{equation}
   i\dot{\epsilon}_n - \omega\epsilon_n + \frac{1}{2h^2}(\epsilon_{-}
   - 2\epsilon_{n} + \epsilon_{+}) +
   2f_n^2\epsilon_n+f_n^2\epsilon_n^*=0. \label{LinearizedClassic}
\end{equation}

Similarly we obtain the linearized equations for $\epsilon_n(t)$
for the $N$-conserving model $II$ of Eq. (\ref{Model2}),
\begin{eqnarray}
   i\dot{\epsilon}_n - \omega\epsilon_n + \frac{1}{2h^2}
   (\epsilon_{-} - 2\epsilon_{n} + \epsilon_{+}) \nonumber \\
   +\frac{1}{2} f_n(f_{-}+f_{+}) \left[ \epsilon_{n}+ {\rm Re}
   (\epsilon_{n}) \right] \nonumber \\+ \frac{1}{2} f_n^2\left[{\rm
   Re} (\epsilon_{-})+{\rm Re} (\epsilon_{+})\right] =0,
   \label{LinearizedII}
\end{eqnarray}
and for the $\tilde{N}$- and $P$-conserving model $III$ of Eq.
(\ref{Model3}),
\begin{eqnarray}
   i\dot{\epsilon}_n - \omega\epsilon_n + \frac{1}{2h^2}
   (\epsilon_{-} - 2\epsilon_{n} + \epsilon_{+}) \nonumber \\
   +\frac{1}{2}(f_{-}+f_{+})\left[f_{-}{\rm Re}
   (\epsilon_{-})+f_{+}{\rm Re}(\epsilon_{+}) \right] \nonumber \\
   +\frac{1}{4} \left(f_{-}^2 + f_{+}^2\right)
   (\epsilon_{-}+\epsilon_{+})=0. \label{LinearizedIII}
\end{eqnarray}

Defining $\epsilon_{n}(t)=a_n(t)+ib_n(t)$, the linearized
equations can be written as $\left( {\begin{array}{*{20}c}
   {{\mathbf{\dot a}}}  \\
   {{\mathbf{\dot b}}}  \\

 \end{array} } \right) = \left( {\begin{array}{*{20}c}
   0 & {\mathbf{\Omega }}  \\
   {\mathbf{J}} & 0  \\

 \end{array} } \right)\left( {\begin{array}{*{20}c}
   {\mathbf{a}}  \\
   {\mathbf{b}}  \\

 \end{array} } \right),$
where vectors $\mathbf{a}$ and $\mathbf{b}$ contain $a_n$ and
$b_n$, respectively. Stationary soliton solutions are linearly
stable if and only if the eigenvalue problem, $\det(\mathbf{J}
\mathbf{\Omega } -\lambda^2 \mathbf{I})$, has only real and
nonpositive solutions for $\lambda^2$ \cite{Carr:1985-201:PLA}.

In our stability analysis, a soliton, having a frequency $\omega=1$,
is placed at the middle of chain of 400 sites with periodic
boundary conditions. Different magnitudes of the discreteness
parameter $h$ and different positions of solitons with respect to
the lattice, $x_0$, are considered.

Most of eigenfrequencies $\sqrt{-\lambda^2}$ appear within the
band between $\omega$ and $\omega+2/h^2$, where $\omega$ is the
frequency of soliton and $2/h^2$ is the maximum frequency of the
linear spectrum of trivial solution, $f_n=0$. We do not show these
eigenvalues in the figures. Eigenvalues appearing outside of this
band are related to soliton's internal modes. The spectra of solitons
in the models $I$, $II$, and $III$ always contain two zeroes.
However, spectra of solitons in the PNp-free models (including
models $II$ and $III$), due to their effective translational
invariance, always contain two additional zeroes.

\begin{figure}
\includegraphics{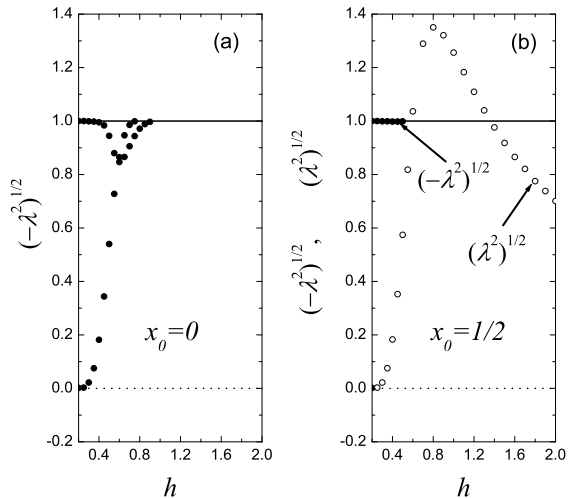}
\caption{Model $I$, spectrum of the soliton with frequency
$\omega=1$ (bottom edge of the continuous spectrum) for (a) stable
on-site and (b) unstable inter-site configurations. Spectrum
always contains a pair of zero-frequency modes shown by dashed
lines. For $h<0.4$ discreteness is weak and the soliton in (a) has
a mode with nearly zero frequency corresponding to the
translational mode of the continuum NLS equation. The frequency of
this mode grows rapidly with increase in $h$ for $h>0.4$ and it
enters the continuous band at $h=0.75$ where soliton loses its
mobility.} \label{Figure1}
\end{figure}

Results for the classical DNLS equation (model $I$) are presented
in Fig. \ref{Figure1} for (a) $x_0=0$ and (b) $x_0=1/2$. In (a),
the on-site soliton is stable because all $\lambda^2$ are real and
nonpositive, while in (b), the inter-site soliton is unstable
because we have positive $\lambda^2$ (shown by open circles).
The solid line shows the bottom edge of the linear band and dotted
line shows the always existing eigenvalues $\lambda^2=0$.

\begin{figure}
\includegraphics{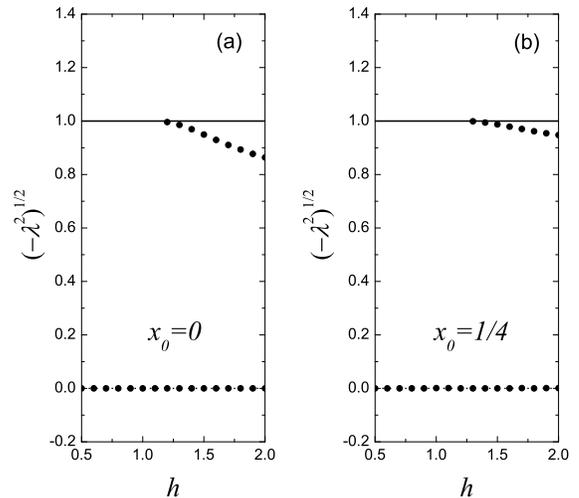}
\caption{Model $II$, spectrum of the soliton with frequency
$\omega=1$ for (a) on-site and (b) asymmetric configurations. Both
configurations are stable. A soliton internal mode bifurcates from
the bottom edge of the continuous frequency band at rather large
$h$. In addition to an always existing pair of zero-frequency modes
(shown by dashed lines), for any $x_0$, there exist another pair
of zero-frequency modes (shown by dots), reflecting the
translational invariance of the soliton and the absence of the
Peierls-Nabarro potential.} \label{Figure2}
\end{figure}

In Fig. \ref{Figure2}, the results for the $N$-conserving,
PNp-free DNLS equation (model $II$) are presented. The always
existing eigenvalues $\lambda^2=0$ are shown by dotted line and
the additional two zeroes are shown by dots; the latter reflect
the translational invariance of the soliton. Note that panel (b)
corresponds to the soliton having $x_0=1/4$, i.e., placed
non-symmetrically with respect to the lattice points. For $h>
1.25$, below the linear spectrum, there exists a soliton internal
mode. This mode was not observed for the soliton placed at
$x_0=1/2$.

\begin{figure}
\includegraphics{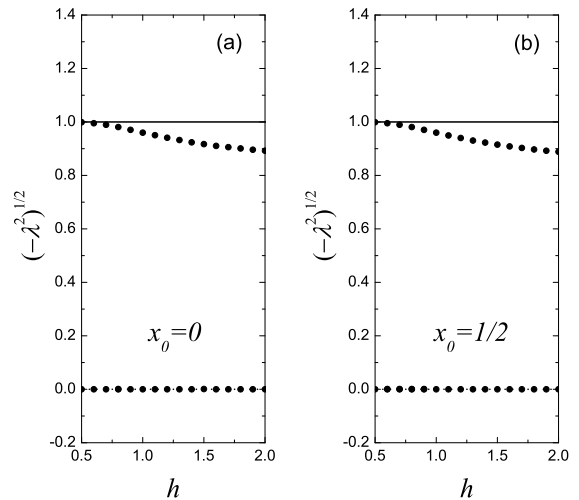}
\caption{Model $III$, bottom part of the spectrum of the soliton
with frequency $\omega=1$ for (a) on-site and (b) inter-site
configurations. Both configurations are stable. Soliton internal
mode bifurcates from the bottom edge of the continuous frequency
band. For any $x_0$, the spectrum contains two pairs of zero-frequency
modes, which is a common feature for the PNp-free models.}
\label{Figure3}
\end{figure}

\begin{figure}
\includegraphics{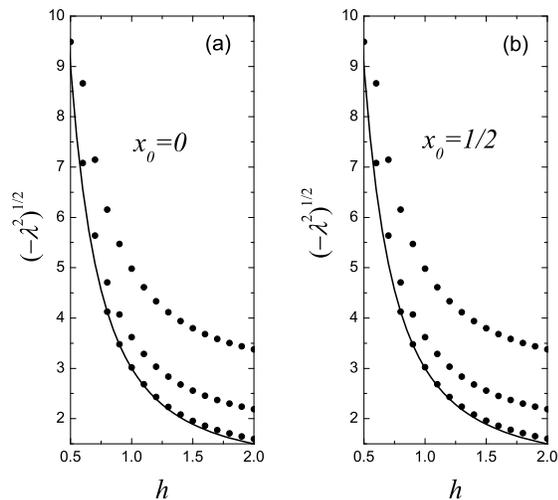}
\caption{Same as in Fig. \ref{Figure3} but for the top part of the
spectrum. In contrast to models $I$ and $II$, the soliton in the model
$III$ has internal modes with frequencies lying above the
continuous band. Solid line shows the upper edge of the continuous
band.} \label{Figure4}
\end{figure}

Similar results for the $\tilde{N}$- and $P$-conserving, PNp-free
DNLS equation (model $III$) are shown in Fig. \ref{Figure3}. In
contrast to models $I$ and $II$, soliton in model $III$ has the
internal modes lying not only below, but also above the linear
band. These modes are shown in Fig. \ref{Figure4} by dots and the
upper edge of the linear spectrum is shown by the solid line.

\subsection{Mobility of solitons}

\begin{figure}
\includegraphics{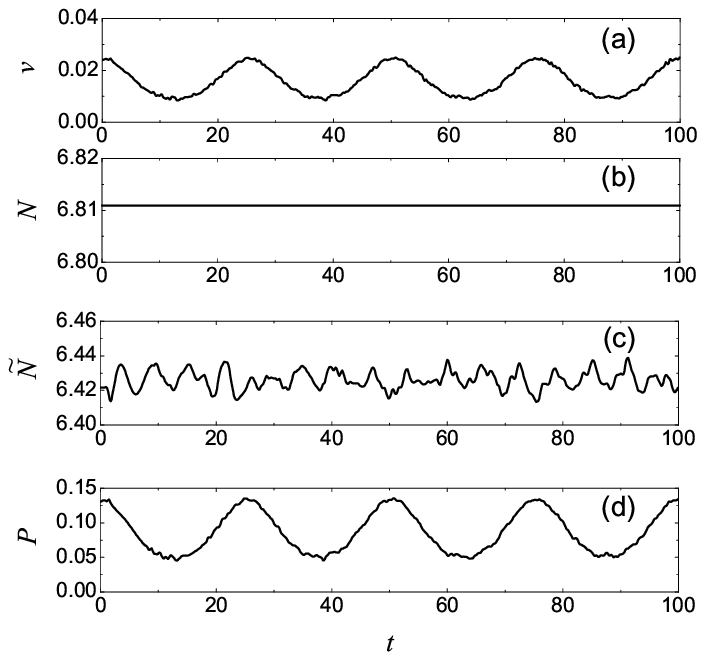}
\caption{Model $I$, $h=0.4$. Time evolution of (a) soliton
velocity, (b) norm, (c) modified norm, and (d) momentum. Soliton
frequency is $\omega=1$.} \label{Figure5}
\end{figure}

\begin{figure}
\includegraphics{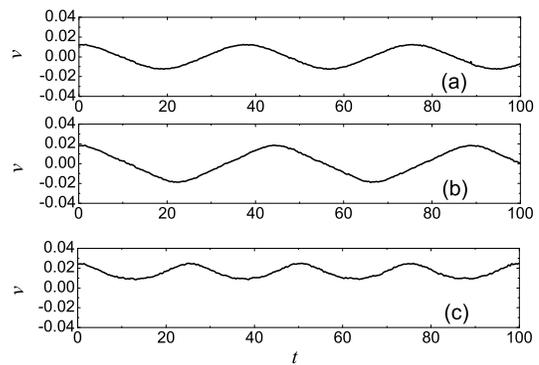}
\caption{Model $I$, $h=0.4$. Time evolution of soliton velocity
for different initial velocities. Soliton frequency is
$\omega=1$.} \label{Figure6}
\end{figure}

\begin{figure}
\includegraphics{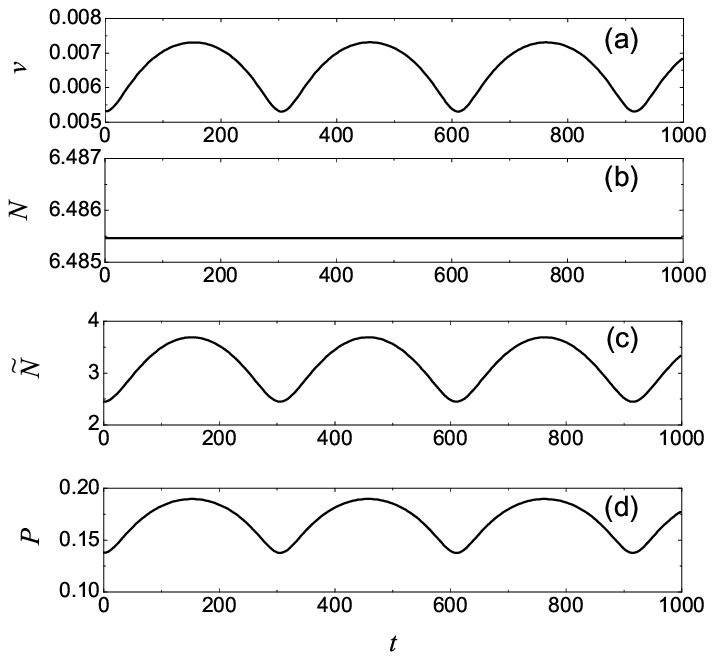}
\caption{Model $II$, $h=2.0$. Time evolution of (a) soliton
velocity, (b) norm, (c) modified norm, and (d) momentum. Soliton
frequency is $\omega=1$.} \label{Figure7}
\end{figure}

\begin{figure}
\includegraphics{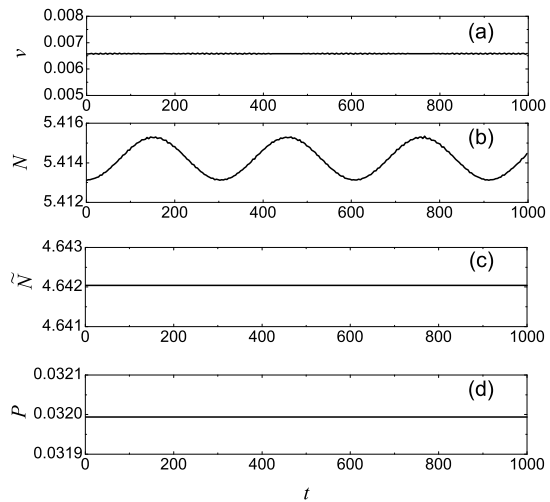}
\caption{Same as in Fig. \ref{Figure7} but for model $III$.}
\label{Figure8}
\end{figure}

Solving the eigenvalue problem formulated in Sec.
\ref{InternalModes}, we find solutions of corresponding DNLS
equation of the form of Eq. (\ref{Perturbation}). Solutions are
accurate for small amplitudes of the eigenvectors, $\epsilon_n(t)$.
Here we examine the solutions corresponding to the translational
eigenmode (since ``kicks'' along this eigendirection may be
responsible for/related to motion in these lattices).

We define the position of discrete soliton as its center of mass:
\begin{equation}
   S(t)=\frac{\sum n N_n(t)}{\sum N_n(t)}, \label{SolitonPosition}
\end{equation}
where $N_n=|\psi_n|^2$; then the soliton velocity is $v=dS/dt$. This
definition is used for $N$-conserving models $I$ and $II$, and for
$\tilde{N}$-conserving model $III$, in Eq.
(\ref{SolitonPosition}), we naturally use
$\tilde{N}_n=(1/4)(\psi_n \psi_{-}^{\star} + \psi_n^{\star}
\psi_{-}+\psi_n \psi_{+}^{\star} + \psi_n^{\star} \psi_{+})$
instead of $N_n$.

Setting initial conditions according to Eq. (\ref{Perturbation})
with sufficiently small amplitude of translational eigenmode
$\epsilon_n(t)$, we then integrate numerically the DNLS equations
to study the soliton mobility at different $h$.

For all three models we boost the solitons initially placed at
$x_0=0$.

As Fig. \ref{Figure1}(a) suggests, the soliton of model $I$ has a
translational mode with nearly zero frequency only for $h<0.4$.
For such a small $h$, discreteness is weak and model $I$ can be
regarded as weakly perturbed continuum NLS equation, which
supports moving solitons. However, the frequency of the translational
mode increases rapidly for $h>0.4$ and it enters the continuum
frequency band at $h=0.75$ when the soliton completely loses its
mobility.

In Fig. \ref{Figure5} we show the time evolution of soliton's
velocity, $v$, norm, $N$, modified norm, $\tilde{N}$, and
momentum, $P$, for the soliton of frequency $\omega=1$ in model
$I$ at $h=0.4$. Soliton was boosted with the initial velocity of
$v=0.024$. One can see that norm is conserved while all the other
parameters oscillate while the soliton propagates along the chain.
The soliton's velocity  is maximum (minimum) when it passes the
on-site (inter-site) configuration with minimum (maximum) energy.
Its average speed is equal to $0.016$.

To boost the soliton in model $I$ one has to apply sufficiently
large initial momentum (i.e., larger than a threshold in order)
to overcome the Peierls-Nabarro potential.
This feature is illustrated by Fig. \ref{Figure6}, where we boost
the soliton with different initial velocities: (a) $v_0=0.012$,
(b) $v_0=0.018$, and (c) $v_0=0.024$ (same as in Fig.
\ref{Figure5}). The soliton does not propagate in (a) and in (b),
instead, it oscillates near the minimum energy configuration.

The soliton kinematics/dynamics in the PNp-free models $II$ and $III$ are
qualitatively different from those in model $I$ because soliton is
not trapped by the lattice and
thus can be accelerated by even weak external fields.

In Fig. \ref{Figure7} and Fig. \ref{Figure8} we show the results
for models $II$ and $III$, respectively, similar to that presented
in Fig. \ref{Figure5} for model $I$. Note that for model $I$ we
used a rather small discreteness parameter $h=0.4$ while solitons in
Fig. \ref{Figure7} and Fig. \ref{Figure8} propagate along
highly-discrete chains at $h=2$. It can be clearly seen that model
$II$ conserves classical norm $N$, while model $III$ conserves
modified norm $\tilde{N}$ and momentum $P$.

\section{Conclusions} \label{sec:Conclusions}

We have described a general and systematic method of constructing
spatial discretizations of NLS-type models, whose stationary
soliton solutions can be obtained from a two-point difference
problem. In this setting, finding stationary solutions becomes
tantamount to solving simple nonlinear algebraic equations. We
have also illustrated the connections of the resulting models with
the integrable discretization of the NLS equation, of which they
are a natural generalization for cubic nonlinearities (our
construction was given for arbitrary polynomial
nonlinearities of a particular parity);
furthermore, the differences of such models from
the standard discretization of the NLS equation often encountered
in physical applications have been highlighted, both in terms of
the relevant dynamical (solitonic) behavior as well as in terms of
the underlying conservation laws present in the various models.

For the case of cubic nonlinearity we demonstrate that in the
constructed PNp-free DNLS chains solitons are stable in a wide
range of the discreteness parameter $h$.
Semi-analytical moving soliton solutions are found in the regime
of propagation at slow speed.
Several numerical examples illustrate the qualitative difference
in slow soliton dynamics in the classical DNLS equation and the
PNp-free DNLS equations. In the classical model to boost the
soliton one has to apply sufficiently large initial momentum to
overcome the Peierls-Nabarro potential. In the PNp-free model,
the soliton behaves differently and can be
accelerated by even weak external fields.

We believe that our results may suggest novel possibilities for engineering nonlinear lattices optimized for efficient control over the localization and propagation of discrete solitons in various physical contexts.

\section*{Acknowledgements}

We would like to acknowledge a number of useful discussions with
Yu. S. Kivshar and also with D.J. Frantzeskakis.
SVD wishes to thank the warm hospitality of the
Nonlinear Physics Centre at the Australian National University.
PGK gratefully acknowledges the support of NSF-DMS-0204585,
NSF-DMS-0505063 and NSF-CAREER.



\end{document}